\documentclass[useAMS,usenatbib,fleqn]{mn2e}

\usepackage{aas_macros}

\usepackage{epsfig}
\usepackage{epstopdf}
\usepackage{subfigure}
\usepackage{amsmath}    
\usepackage{longtable}
\usepackage{dcolumn}
\usepackage{amssymb}

\usepackage{upgreek} 

\usepackage{wrapfig}

\DeclareGraphicsExtensions{.jpg,.eps,.EPS,.png,.pdf,.ps}

\usepackage{colordvi}
\usepackage{ulem}

\RequirePackage[colorlinks=true
,urlcolor=blue
,anchorcolor=blue
,citecolor=blue
,filecolor=blue
,linkcolor=blue
,menucolor=blue
,pagecolor=blue
,linktocpage=true
,pdfproducer=medialab
]{hyperref}

\usepackage[usenames,dvipsnames]{xcolor}

\title[Associated 21-cm FRB Absorption]{Inferring the Distances of Fast Radio Bursts Through Associated 21-cm Absorption}

\author[B. Margalit \& A. Loeb]
{Ben Margalit$^{1,2}$\thanks{E-mail: \href{mailto:btm2134@columbia.edu}{btm2134@columbia.edu}} and Abraham Loeb$^{2}$ \\ 
\normalsize $^{1}$Physics Department, Columbia University, 538 West 120th St., New York, NY 10027 \\
\normalsize $^{2}$Institute for Theory and Computation, Harvard University, 60 Garden St., Cambridge, MA 02138, USA \\
}

\date{}

\begin{document}
\maketitle

\begin{abstract}
The distances of Fast Radio Burst (FRB) sources are currently unknown. We show that the 21-cm absorption line of hydrogen can be used to infer the redshifts of FRB sources, and determine whether they are Galactic or extragalactic. We calculate a probability of $\sim 10\%$ for the host galaxy of an FRB to exhibit a 21-cm absorption feature of equivalent width $\gtrsim 10 ~\mathrm{km}~\mathrm{s}^{-1}$. Arecibo, along with several future radio observatories, should be capable of detecting such associated 21-cm absorption signals for strong bursts of $\gtrsim {\rm several}~\mathrm{Jy}$ peak flux densities.
\end{abstract}


\section{Introduction}
Fast radio bursts (FRBs), a recently discovered class of transient events in which a $\sim 1 ~\mathrm{Jy}$ signal of duration $\sim 1 ~\mathrm{ms}$ is observed at radio frequencies of $\sim 1-10~\mathrm{GHz}$, have become the source of great debate in the astrophysical community. The large dispersion measures (DM) of these events have led to the suggestion of their extragalactic origin, implying high isotropic luminosities of $\sim 10^{42} ~\mathrm{erg}~\mathrm{s}^{-1}$. Despite an estimated rate of order $\sim 10^3 ~\mathrm{day}^{-1}~(4\uppi ~\mathrm{sr})^{-1}$ \citep{Rane+2015}, only a handful of events have been reported to date. Of these, twelve were found in analysis of archival data from the Parkes radio telescope \citep{Lorimer+2007,Keane+2012,Thornton+2013,BurkeSpolaor+2014,Champion+15}, one in archival data from the Arecibo radio telescope \citep[FRB121102;][]{Spitler+2014}, and another in archival data of the Green Bank Telescope \citep[GBT;][]{Masui+15}. Three FRBs have been discovered in real-time at the Parkes radio telescope \citep{Ravi+2015,Petroff+2015,Keane+16}. \cite{Keane+16} claimed to have identified an associated FRB `afterglow' which would have provided the first FRB redshift measurement, but this claim has largely been refuted by \cite{Williams&Berger16} and later \cite{Vedantham+16} \citep[although see][]{Li&Zhang16}. In an important discovery, \cite{Spitler+16} have recently reported repetitions following FRB121102, later confirmed with additional repetitions found by \cite{Scholz+16}.

Although many theoretical models for these events had been proposed, the physical mechanism producing FRBs as well as their origin remains illusive. Most models assume an extragalactic origin in the possible contexts of black-hole evaporation \citep{Keane+2012}, magnetar hyperflares \citep{PoPov&Postnov2013,Katz2014}, neutron star mergers \citep{Totani2013,Ravi&Lasky2014}, white dwarf mergers \citep{Kashiyama2013}, collapse of supramassive neutron stars \citep{Falcke&Rezzolla2014,Zhang2014}, orbiting bodies immersed in pulsar winds \citep{Mottez&Zarka2014}, magnetar pulse-wind interactions \citep{Lyubarsky2014}, giant pulses from magnetars near galactic centers \citep{Pen&Connor2015}, collisions between neutron stars and asteroids \citep{Geng&Huang2015}, and giant pulses from young pulsars within supernova remnants \citep{Connor+2015,Katz2015}.
Alternatively, \cite{Loeb+2014} suggested that FRBs may originate from nearby flaring stars in our own Galaxy \citep[see also][]{Maoz+2015}, and \cite{Kulkarni+2014}, motivated by discoveries of terrestrial FRB-like impostors \citep{Burke-Spolaor+2011}, suggested a terrestrial origin for FRBs.
Note that a terrestrial origin has largely been ruled out due to FRB121102's repetitions, which have been detected at both Arecibo and GBT with consistent localization and DM. These repetitions are also in strong tension with catastrophic event models (usually extragalactic), unless they form a sub-population of FRBs.

In this paper, we propose a novel method for inferring the redshift of FRBs. This method could allow precise distance measurements independent of the DM, and may therefore break the degeneracy between terrestrial, Galactic and extragalactic interpretations of the large electron column densities. 
Specifically, we suggest measuring the absorption signature of the 21-cm line of neutral hydrogen in the FRB's host galaxy. Detection of an absorption line at longer wavelengths $\lambda_\mathrm{obs}$ yields an estimated measurement of the FRB progenitor's redshift, $z = (\lambda_\mathrm{obs}/21.106~\mathrm{cm}) - 1$. The precision of this method is not limited by the unknown electron column density along line-of-sight, in sharp contrast to DM estimates of the distance scale. 

Recently, \cite{Fender&Oosterloo2015} have also considered the signature of the 21-cm line, but focused on absorption by intervening intergalactic clouds (as well as absorption in the Milky Way) whereas we discuss absorption in the FRB's host galaxy. 
These studies are complementary in this sense. Our estimates indicate that associated absorption is more likely to yield high HI column densities. This is consistent with the fact that only a small fraction of quasars show evidence for damped Ly$\alpha$ absorption which flags galactic disks \citep[][and references therein]{Fumagalli+2014}. Even so, a redshift inferred from 21-cm absorption should conservatively be interpreted as only a lower-bound on the true FRB redshift due to the possibility of intervening absorption.

Successful application of our method relies on HI absorption features being strong and common among FRBs, which in turn depend on the HI column density traversed by a typical FRB. In the following sections we quantify the associated statistics. 
Throughout our analysis, we assume that FRBs originate from common gas-rich galaxies, and not from rare galaxies which are deficient in cold gas and would likely not exhibit any HI absorption.
Our results for 21-cm absorption are presented in \S \ref{sec:21cm_Absorption}. We continue by estimating the probability for significant HI absorption in FRBs (\S \ref{sec:Probability}). Finally, \S \ref{sec:Discussion} summaries our results and their application to current and future radio telescope surveys.

\section{21-cm Absorption} \label{sec:21cm_Absorption}
The optical depth to 21-cm absorption of HI gas is given by \citep[e.g][]{Loeb2008},
\begin{equation} \label{eq:tau_nu}
\tau (\nu) = \frac{3 A_{10} h c^2}{32 \uppi \nu_{10}} \frac{N_\mathrm{HI}}{k_B T_\mathrm{s}} \phi(\nu) ,
\end{equation}
where $A_{10}\approx 2.8689 \times 10^{-15} ~\mathrm{s}^{-1}$ is the Einstein coefficient for the hyperfine transition, $\nu_{10} \approx 1.4204 ~\mathrm{GHz}$ is the frequency associated with the transition energy, $\phi(\nu)$ is the line profile function (normalized such that $\int \phi \ d\nu = 1$), $N_\mathrm{HI}$ is the intervening neutral hydrogen column density, and $T_\mathrm{s}$ is the spin temperature.

It is commonly assumed that the spin temperature is in equilibrium with the kinetic temperature of the HI gas, since collisions dominate the level population of the hyperfine transition at the gas densities of interest.
The interstellar medium (ISM) is known to exhibit multiple phases of neutral hydrogen in pressure equilibrium: the cold neutral medium (CNM) whose characteristic temperature is $\sim 100~\mathrm{K}$, and the warm neutral medium (WNM) which is typically at $\sim 5,000~\mathrm{K}$. In order to maintain pressure equilibrium, the CNM must be significantly denser than the WNM, however it is also observed to have a lower volume filling factor. Coincidentally, these opposing factors conspire to give very similar average densities $f_\mathrm{CNM} n_\mathrm{CNM} \sim f_\mathrm{WNM} n_\mathrm{WNM} \sim 0.1 ~\mathrm{cm}^{-3}$ \citep[e.g.][]{Draine2011}.
It therefore follows that the main contribution to HI absorption originates from the CNM. This is because the average column densities $N_\mathrm{HI} = \int f n_\mathrm{HI} \ ds$ through either the CNM or the WNM are of the same order of magnitude, whereas the $T^{-1}$ falloff in Eq.\ (\ref{eq:tau_nu}) favors the lower temperature CNM.

Defining the integrated optical depth as $\tau \Delta v \equiv \int \tau(v) \ dv$,  we find
\begin{equation} \label{eq:tau_Delta_v}
\tau \Delta v \approx 0.55 ~\mathrm{km}~\mathrm{s}^{-1} ~\left(\frac{N_\mathrm{HI}}{10^{20}~\mathrm{cm}^{-2}}\right) ~\left(\frac{T_\mathrm{s}}{100 ~\mathrm{K}}\right)^{-1} .
\end{equation}
This quantity provides a measure of the total attenuation over all frequency bands, and is commonly referred to as the `equivalent width' of the line. Once the characteristic line width $\Delta v$ is prescribed, it is straightforward to calculate the typical optical depth $\tau$ from Eq.\ (\ref{eq:tau_Delta_v}). Viewed this way, $\tau \Delta v$ can be thought of as a top-hat approximation for the integral $\int \tau(v) \ dv$.

The intrinsic hyperfine line width is extremely narrow, so that in any practical application the minimal observed line width is set by the thermal, turbulent, or bulk velocities of the gas. 
For a thin Milky-Way-like disk, the turbulent velocity is typically $\sigma_t \sim 10~\mathrm{km}~\mathrm{s}^{-1}$ even though within a single CNM cloud at $T=100~\mathrm{K}$, the sound speed is only $\sim 1~\mathrm{km}~\mathrm{s}^{-1}$. 
Additional broadening from large scale motions of the gas, such as galactic rotation, depends on the inclination angle of the host galaxy relative to the line-of-sight. Since the typical distance propagated through the host galaxy disk is of order the vertical scale-height $z_d$, this effect results in $\Delta v \sim (z_d / R_d) v_\mathrm{rot} \approx \sigma_t$ as well. Here, $R_d$ refers to the disk radial scale-length, and we have made use of the well known thin disk relation (obtained from vertical hydrostatic balance) $z_d / r \approx \sigma / v_\mathrm{rot}$ \citep[e.g.][]{Frank&King&Raine2002}. Note that the mean distance traveled through the galaxy is $2z_d$ and that in rare cases if the disk is viewed nearly edge on, $\Delta v$ can be substantially larger due to galactic rotation.

The value of $\tau \Delta v$ in Eq.\ (\ref{eq:tau_Delta_v}) is characteristic of a Milky Way like galaxy, since a reasonable estimate for the HI column density is $N_\mathrm{HI} \sim n_\mathrm{HI} z_d \sim 10^{20} ~\mathrm{cm}^{-2}$. Even so, significantly smaller or larger value are possible depending on the FRB source's location within its host galaxy, the distribution of neutral hydrogen in the galaxy, the inclination angle, and the host disk properties. In the following section we calculate the probability of measuring various values of $\tau \Delta v$, taking these factors into account.

\section{Probability Distribution of HI Absorption Signals}  
\label{sec:Probability}
In estimating the probability distribution for finding various $\tau \Delta v$ values, we assume a fixed CNM spin temperature of $T_\mathrm{s} = 100~\mathrm{K}$. Calculating $\tau \Delta v$ then reduces to calculating the HI column density, which we split into two separate components, 
$N_\mathrm{HI} = N_0(M_\mathrm{HI}) \eta({\bf x_0}, i,\varphi)$.
Here
\begin{equation} \label{eq:N_0_Definition}
N_0(M_\mathrm{HI}) \equiv \frac{M_\mathrm{HI} / m_\mathrm{H}}{4\uppi R_d(M_\mathrm{HI})^2} \sim n_\mathrm{HI} z_d \ ,
\end{equation}
is a characteristic column density for a host galaxy containing a mass $M_\mathrm{HI}$ in neutral hydrogen. The second term, $\eta({\bf x_0}, i,\varphi)$, is a dimensionless geometrical factor which depends on the position of the FRB source inside its host galaxy, ${\bf x_0}$, and on the inclination and azimuthal angles $i$, $\varphi$ relative to the line-of-sight.

\begin{figure}
\epsfig{file=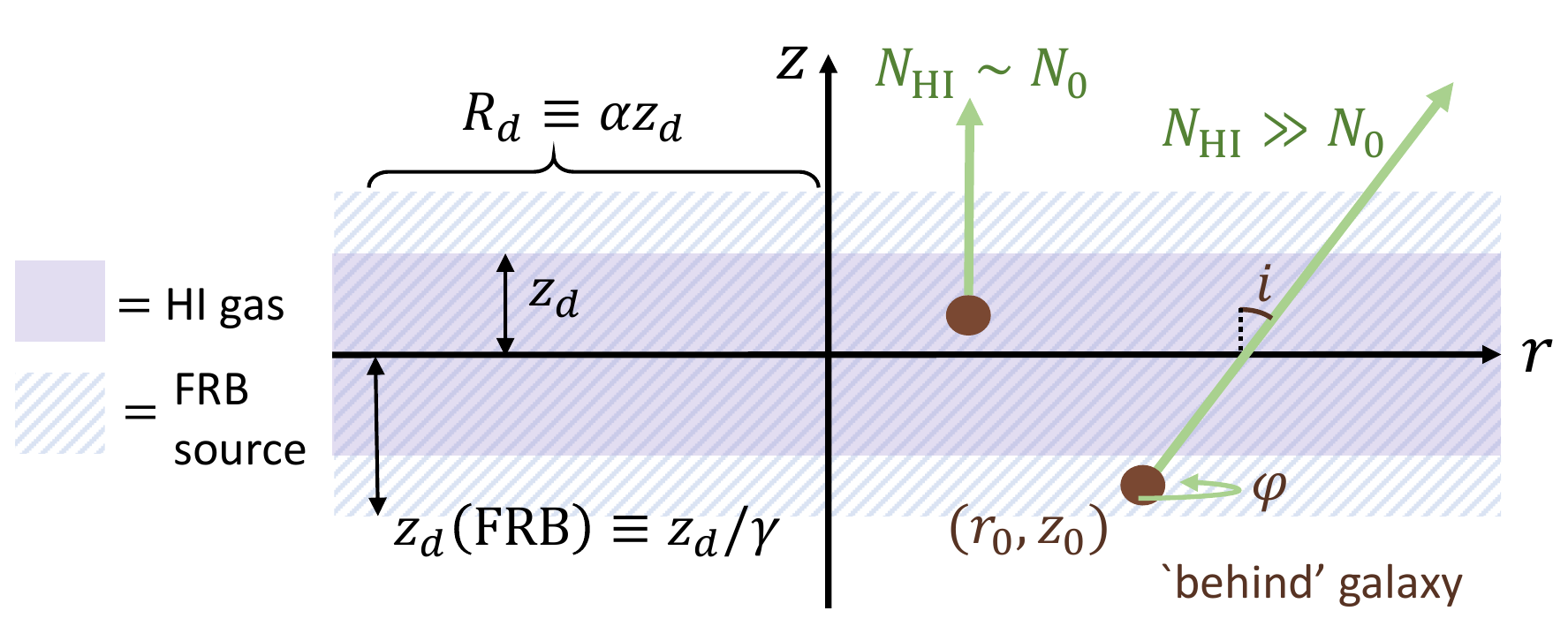,angle=0,width=0.45\textwidth}
\caption{Schematic diagram of a galactic disk (cross section) illustraing the geometry of our model. The lightly shaded purple region demarcates the neutral hydrogen whereas the hatched light blue area shows the FRB source distribution. Our exponential disk model assumes that both components are characterized by the same radial scale-length, $R_d$. The FRB population's vertical scale-height can vary by a factor of $1/\gamma$ relative to the HI scale-height, $z_d$. The brown circles are two example FRB progenitors, and the long green arrows indicate the line-of-sight towards the observer.} \label{fig:Disk_Sketch}
\end{figure}

We consider the neutral hydrogen gas and FRB source distributions to be exponential disks. A schematic diagram illustrating our model geometry and key parameters is given in Fig.\ \ref{fig:Disk_Sketch}. We take the radial scale-length of the disks, $R_d$, to be the same for both components, yet allow the FRB source population's vertical scale-height to vary by a factor of $1/\gamma$ relative to the HI vertical scale-height $z_d$, namely
\begin{equation} \label{eq:galactic_distribution}
n \propto \exp \left(-\frac{r}{R_d}\right) \times 
\begin{cases}
\exp \left({-|z|/z_d}\right) , &\mathrm{HI} \\
\exp \left({-\gamma|z|/z_d}\right) , &\mathrm{FRB} \ .
\end{cases}
\end{equation}
Under this prescription, the geometrical factor $\eta$ can be expressed as
\begin{align} \label{eq:eta}
&\eta(r_0,z_0, i, \varphi) \equiv \int n_\mathrm{HI} \ ds \big/ N_0  \\ \nonumber
&= \alpha \int_0^\infty \exp \bigg( -\sqrt{\tilde{r}_0^2 + \tilde{s}^2 \sin^2 i + 2 \tilde{r}_0 \tilde{s} \sin i \cos \varphi} \\ \nonumber 
& ~~~~~~~~~~ - \alpha \left\vert \tilde{z}_0 + \tilde{s} \cos i \right\vert \bigg) \ d\tilde{s} \ ,
\end{align}
where $\tilde{r_0}$ ($\tilde{z_0}$) is defined as $r_0/R_d$ ($z_0/R_d$) and $\alpha \equiv R_d/z_d$ is the ratio of radial to vertical scale-lengths and is roughly $\sim 15$ for the Milky Way.

The cumulative probability distribution for measuring a value of $\eta$ larger than $\eta^\prime$, $P(\eta > \eta^\prime)$, can then be calculated by populating a grid in $(r_0,z_0,i,\varphi)$. Using Eq.\ (\ref{eq:eta}), each gridpoint can be assigned a well defined value of $\eta$, as well as a probability density for FRB sources to populate its region, $f({\bf x_0},i,\varphi) \propto n_\mathrm{FRB}(r_0,z_0) r_0 \sin i$. Generating a sufficiently extended and finely coarsed grid, we sort the gridpoints in decreasing order of $\eta$ and cumulatively sum the probabilities $dP \approx f(r_0,z_0,i) dr_0 dz_0 di d\varphi$. This results in a numerical evaluation of $P(\eta>\eta^\prime)$ which is equivalent to solving the integral $\int dr_0 dz_0 di d\varphi \ f(r_0,z_0,i) \Theta \left[\eta > \eta^\prime(r_0,z_0,i)\right]$, where $\Theta$ is the Heaviside function.

Once $P(>\eta)$ is obtained, all that remains in the process of calculating $P(>\tau \Delta v)$ is an estimation of $N_0$. Several studies have shown that the HI mass component of spiral galaxies scales as the disk scale-radius squared, namely $M_\mathrm{HI} \propto R_d^2$ \citep[e.g.][]{Broeils&Rhee1997,Verheijen&Sancisi2001}, so that the surface density $N_0$ is essentially constant and does not vary appreciably with galactic mass. 

We calculate the characteristic (mean) column density $\left\langle N_0 \right\rangle$ using the tabulated HI masses and radii of \cite{Broeils&Rhee1997}. We relate our exponential disk scale-length, $R_d$, to the radius $R_{\rm eff}$ which encloses half the HI mass and is quoted by these authors. For our exponential disk model~(\ref{eq:galactic_distribution}), we find that the radius encompassing 50\% of the total HI mass is $R_{\rm eff} \approx 1.68 R_d$. 
Substituting this relation in Eq.~(\ref{eq:N_0_Definition}) and using values of $M_{\rm HI}$ and $R_{\rm eff}$ from table~1 of \cite{Broeils&Rhee1997}, we find that
the resulting mean column density and its standard deviation are
\begin{equation} \label{eq:N_0}
\langle N_0 \rangle = \left( 9.1 \pm 3.6 \right) \times10^{20} ~\mathrm{cm}^{-2} ~.
\end{equation}

Note that a crude estimate of this result may be obtained
without the full tabulated data of \cite{Broeils&Rhee1997},
 by using the mean `hybrid' density found by these authors, $\langle \sigma^*_{\rm HI} \rangle = 11.2 ~M_\odot~{\rm pc}^{-2}$. Here the `hybrid' density is defined as $\sigma^*_{\rm HI} \equiv M_{\rm HI} / \uppi R_{25}^2$, and $R_{25}$ is the {\it optical} radius at an isophote of $25~{\rm mag}~{\rm arcsec}^{-2}$. \cite{Broeils&Rhee1997} find that the optical radius strongly correlates with the HI half mass radius $R_{\rm eff}$, $\langle R_{25} / R_{\rm eff} \rangle \approx 1$, which combined with the relation $R_{\rm eff} \approx 1.68 R_d$ and the factor four difference in definition between $\sigma^*_{\rm HI}$ and $N_0$, yields an estimate of $\langle N_0 \rangle \approx 0.7 \langle \sigma^*_{\rm HI} \rangle$.

\begin{figure}
\epsfig{file=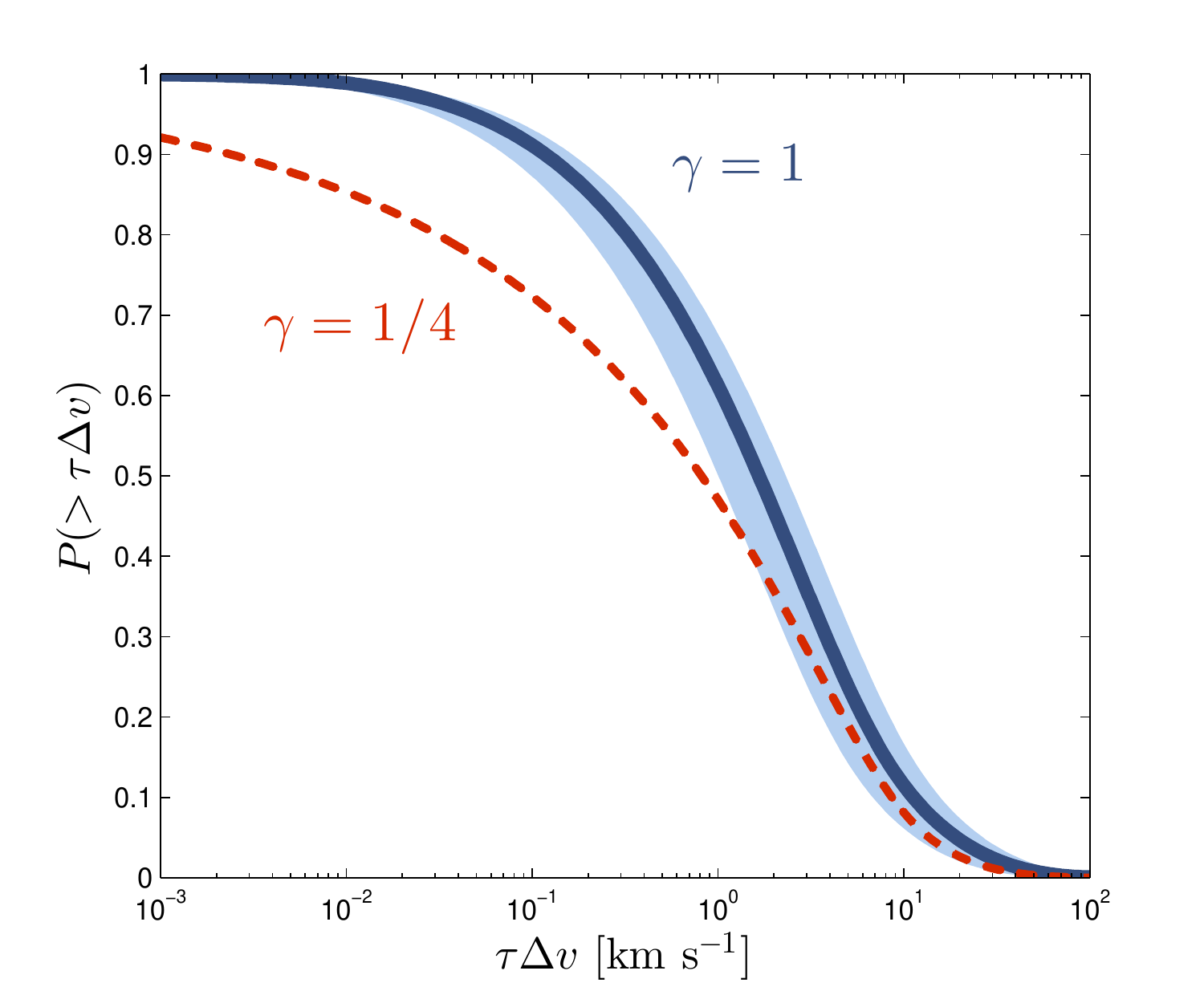,angle=0,width=0.5\textwidth}
\caption{Cumulative probability distribution for measuring an HI absorption feature with equivalent width larger than $\tau \Delta v$, $P(>\tau \Delta v)$. The solid blue curve represents the resulting probability distribution for our canonical values: $\alpha=15$, $\gamma=1$, $T_\mathrm{s}=100~\mathrm{K}$. These values are characteristic of a CNM and a thin disk FRB population. The lightly shaded blue region illustrates the variability due to the $\sim 40\%$ uncertainty in Eq. (\ref{eq:N_0}). This should not be strictly interpreted as the total uncertainty in our results, as the systematics of our model assumptions will likely contribute significantly. The dashed red curve shows the result when the ratio of HI to FRB vertical scale-height is changed to $\gamma=1/4$, characteristic of a thick disk FRB population. As expected, the larger vertical spread in the FRB source distribution reduces the average HI column density and decreases the typical $\tau \Delta v$. The horizontal axis trivially scales as $\left(T_\mathrm{s} / 100~\mathrm{K}\right)^{-1}$ and/or $\left(N_0 / 9.1\times10^{20} ~\mathrm{cm}^{-2}\right)$ if different spin temperatures and/or characteristic column densities are assumed (see Eq.\ \ref{eq:tau_Delta_v}).} \label{fig:P_tau_Delta_v}
\end{figure}

The cumulative probability distribution for measuring HI column density larger than $N^\prime$ is in this case simply related to the probability distribution of $\eta$,
\begin{equation} \label{eq:P_N}
P(N > N^\prime) = P \left[ \eta > N^\prime \big/ N_0 \right] .
\end{equation}
Combining Eq.\ (\ref{eq:tau_Delta_v}), and (\ref{eq:P_N}), we numerically evaluate the probability distribution for $\tau \Delta v$. The results are illustrated in Fig.\ \ref{fig:P_tau_Delta_v}. The thick blue curve shows the cumulative probability distribution for a nominal value of $\gamma=1$, whereas the dashed red curve represents the results for $\gamma=1/4$. For our canonical values ($\gamma=1$), we obtain a median $\tau \Delta v$ of $\approx 1.7 ~\mathrm{km}~\mathrm{s}^{-1}$, and a $10\%$ probability of finding values larger than $\approx 11~\mathrm{km}~\mathrm{s}^{-1}$. Since the parameter $\gamma$ is defined as the ratio of the neutral hydrogen vertical scale-height to that of the FRB source population, the canonical $\gamma=1$ scenario describes an FRB population that is localized to the thin disk component of the galaxy, similarly to the HI distribution.

Figure \ref{fig:P_tau_Delta_v} also shows the resulting cumulative probability distribution for $\gamma=1/4$. This value of $\gamma$ is expected if the FRB source distribution is predominantly associated with the thick disk component of the galaxy, so that the FRB vertical scale-height is larger than the scale-height of the HI disk. As expected, the probability distribution curve in this case is shifted to lower $\tau \Delta v$ values, since there is a larger chance for an FRB to originate further away from the galactic plane where the HI column density along the line of path would (if the line-of-sight is pointed away from the galactic plane) decrease. The median value of $\tau \Delta v$ in this case drops to $\approx 0.8 ~\mathrm{km}~\mathrm{s}^{-1}$, although there is still a $\approx 10\%$ chance of measuring values larger than $9 ~\mathrm{km}~\mathrm{s}^{-1}$.
 
For high values of $\tau \Delta v$, the $\gamma=1/4$ curve converges to the canonical $\gamma=1$ curve, since the largest values of $\tau \Delta v$ are obtained when the FRB progenitor is positioned `behind' its host galaxy. When viewed from the FRB source, the signal propagates towards the galactic plane, passes through the midplane, and continues outwards towards the observer (see Fig.\ \ref{fig:Disk_Sketch}). These large column density events are nearly unaffected by the vertical scale-height of FRB progenitors, hence our predictions for $P(\gtrsim 10 ~{\rm km~s}^{-1})$ are robust with respect to this parameter. We have additionally calculated the probability distributions obtained for various values of $\alpha$ (the ratio of radial to vertical scale lengths) in the range $10-20$, yet find no appreciable change in the results.

\section{Discussion} \label{sec:Discussion}
We have proposed a novel method for measuring the cosmological distances to FRBs, based on the 21-cm absorption signature of HI gas in the FRB progenitor's host galaxy. The method is only useful if the absorption feature is strong enough to be detected. We therefore estimated the probability of obtaining strong HI absorption to asses the feasibility of using the method with current and near future observational facilities.

Using a simple exponential disk model for both FRB and HI galactic distributions  we find a probability of $\sim 10\%$ for measuring values of $\tau \Delta v$ larger than $10~\mathrm{km}~\mathrm{s}^{-1}$, and a median value of $\sim 1 ~\mathrm{km}~\mathrm{s}^{-1}$.

The observational significance of these values can be quantified in terms of the signal-to-noise ratio (SNR) and the channel bandwidth of the radio telescope. If the observational frequency bandwidth (measured in velocity units), $\Delta v_\mathrm{obs}$, exceeds the absorption feature's inherent $\Delta v$, the absorption profile will be unresolved. This, however, does not mean that the HI absorption will be undetectable, since the entire signal in the bandwidth bin will merely be attenuated by a factor of $\approx \tau \Delta v / \Delta v_\mathrm{obs}$. Smaller bandwidth is therefore highly desirable in detecting HI absorption (and note that the previous expression is only correct for $\Delta v_\mathrm{obs} > \Delta v$).

As a feasibility demonstration we focus on the {\it Arecibo Observatory}
\footnote{http://www.naic.edu/}. The sensitivity of this telescope is estimated 
as $A_\mathrm{eff} / T_\mathrm{sys} \approx 1150 ~\mathrm{m}^2~\mathrm{K}^{-1}$, 
where $A_\mathrm{eff}$ is the effective collecting area of the instrument and $T_\mathrm{sys}$ the system temperature. The L-Wide receiver covers the frequency 
range $1.15-1.73~\mathrm{GHz}$, corresponding to a maximum HI line redshift of $\sim 0.24$, whereas the Arecibo L-band Feed Array (ALFA) covers $1.225-1.525~\mathrm{GHz}$ corresponding to $z \leq 0.16$. These are consistent with the low end 
of DM inferred redshifts for known FRBs. With these parameters, and using the 
radiometer equation \citep[e.g.][ch. 12.7]{LoebFurlanetto2013}, we estimate the 
SNR of an FRB with pulse duration $\Delta t$ and peak-flux $F_\nu$ in a frequency 
bin of bandwidth $\Delta v_\mathrm{obs}$ centered around $\nu_\mathrm{obs}$,
\begin{align} \label{eq:SNR}
\mathrm{SNR} &\approx \frac{F_\nu A_\mathrm{eff}}{2 k_B T_\mathrm{sys}} \sqrt{\Delta \nu_\mathrm{obs} \Delta t}
\\ \nonumber
& \approx 4.6 \left( \frac{F_\nu}{1~\mathrm{Jy}} \right) \left( \frac{\Delta t}{3~\mathrm{ms}} \right)^{1/2} 
\left( \frac{\Delta v_\mathrm{obs}}{10~\mathrm{km}~\mathrm{s}^{-1}} \right)^{1/2} \left(\frac{\nu_\mathrm{obs}}{1.2~\mathrm{GHz}}\right)^{1/2} .
\end{align} 
Only for a strong burst would the sensitivity implied by Eq.\ (\ref{eq:SNR}) be sufficient to detect an HI absorption signature. Based on our analysis, and assuming a bandwidth of $\Delta v_\mathrm{obs} = 10~\mathrm{km}~\mathrm{s}^{-1}$, we find a $\sim 10\%$ probability of measuring effective HI optical depths larger than $\sim 1$. These would be detectable at an SNR of $8$ for an FRB of $F_\nu \gtrsim 2~\mathrm{Jy}$. This optimistic assessment does not take into account difficulties associated with noise-confusion, which may be substantial for a single-frequency channel `detection'. Even with such difficulties, a strong burst such as FRB010724 \citep{Lorimer+2007} which had an estimated flux density of $30 \pm 10 ~\mathrm{Jy}$ would likely overcome this issue, and is an ideal candidate for detecting HI absorption.
We additionally note that the repeating nature of at least one FRB \citep{Spitler+16,Scholz+16} could greatly help reduce such noise-confusion and increase a detected absorption feature's significance by implementing stacking techniques (although the large variability in spectral index between repeating bursts poses a clear difficulty).

We also assessed the feasibility of applying the method with future radio observatories such as the {\it Canadian Hydrogen Intensity Mapping Experiment} (CHIME) telescope \citep{Bandura+2014}, the {\it Five hundred meter Aperture Spherical Telescope} (FAST)\footnote{http://fast.bao.ac.cn/en/}, and the {\it Square Kilometer Array} (SKA) telescope\footnote{https://www.skatelescope.org}. 
We repeated the calculation for these observatories and found similar results for FAST and SKA, since the planned sensitivities are roughly comparable with Arecibo: 
$A_\mathrm{eff} / T_\mathrm{sys} \approx 1250, 1630  ~\mathrm{m}^2~\mathrm{K}^{-1}$ for FAST and SKA-mid, respectively\footnote{see e.g. Table 1 of https://www.skatelescope.org/?attachment\_ id=5400}. 
SKA can be an exception to this case, if it is ever realized with a filling factor significantly larger than a few percent.

Our analysis in \S \ref{sec:Probability} is somewhat simplified. To begin with, an exponential disk model with radially fixed vertical scale-height is only a crude model for Milky-Way-like spiral galaxies. Secondly, we have assumed diffuse HI distributions as opposed to more clumpy cloud features representative of the CNM. 
Furthermore, we neglected any redshift evolution in $N_0$ and based our results on studies of galaxies at $z \approx 0$. This last assumption is conservative, since $M_\mathrm{HI}(z)>M_\mathrm{HI}(0)$ and $R_d(z)<R_d(0)$, so that we have underestimated the typical galactic column densities.
We also ignored the possibility that some FRBs originate in elliptical galaxies. As such, our analysis of $P(>\tau \Delta v)$ should be treated as a first crude estimate of the viability of our proposed method.

Scattering effects may also play a role in determining the observed absorption signal and should generally be considered. Diffractive scintillations caused by the intervening ISM or intergalactic medium can, in principle, smear the spectral feature of the 21-cm absorption. Although HI absorption by galactic disks has been successfully observed for quasars \citep[e.g.][]{Borthakur2011}, the potentially smaller source size of FRBs warrants further analysis of the effects of scintillations on their absorption signal.

Finally, we note that while our method provides a means of measuring the FRB's precise redshift, observationally, this scenario is mostly indistinguishable from a possible intervening absorption, which only provides a lower limit on the redshift.
\cite{Fender&Oosterloo2015} estimate a probability of significant intervening absorption ($\tau \Delta v > 2~{\rm km ~s}^{-1}$) at $z=2$ of $\sim 10\%$, but scaling this to typical DM inferred FRB redshifts of $z \sim 0.5$ reduces the probability to $\lesssim 2\%$. This is significantly smaller than the probabilities estimated in our present work, and consistent with the fact that only a small fraction of all quasars show damped Lyman-alpha absorption in their spectra, although caution is necessary in interpreting this result due to the many uncertainties in both models. Further detailed analysis of both associated and intervening absorption is therefore necessary to more accurately asses the relative importance of the two effects.

It is important to stress that the limitations of our analysis affect only the probability distributions for measuring $\tau \Delta v$. This information is important in estimating the likelihood of a 21-cm absorption detection, but bears no implications whatsoever to the application of our method once such an absorption signal is detected. In particular, precise redshift measurements to within $\sim (\Delta v_\mathrm{obs} / c)$ accuracy (neglecting peculiar motions of the host galaxies) are immediately possible following an HI absorption signature detection, regardless of the probability in making such a detection.

While an HI absorption detection would directly confirm the extragalactic origin of FRBs, a non-detection does not exclude this possibility, as a variety of reasons may cause a lack in HI absorption features. For example, it is possible to posit that FRBs originate from elliptical galaxies which typically have smaller HI column densities \citep{Serra+2012}. Alternatively, even if FRBs are assumed to originate in spiral galaxies, there exists a non-zero probability for obtaining weak absorption features that could go undetected. In order to state with any statistical significance that absorption features are systematically lacking would require a large sample of high SNR FRBs, and a more careful analysis of $P(>\tau \Delta  v)$.

As a final note, we point out that Milky Way HI absorption at $1.42 ~\mathrm{GHz}$ should be present in addition to the extragalactic redshifted absorption feature which we have studied. This additional absorption component can be evaluated based on existing data for the Milky Way HI distribution at the measured Galactic latitude and longitude $(l,b)$ towards the FRB source, and is explored in greater detail in \cite{Fender&Oosterloo2015}. An HI column along the FRB direction that is smaller than the total value of the Milky Way disk in that direction would indicate a local Galactic origin for the FRB, in which case the DM must be intrinsic to the immediate environment of the FRB source \citep{Loeb+2014,Maoz+2015}.

\section*{Acknowledgements}
We thank Vicky Kaspi and Michael Johnson for helpful discussions, and the anonymous referee for comments which greatly helped to improve a previous version of this paper. This work was supported in part by NSF grant AST-1312034. B.M. is delighted to thank Brian Metzger for his support and guidance, and the Institute for Theory and Computation for its hospitality.

\bibliographystyle{mn2e}
\bibliography{FRB_Bibliography}

\end{document}